\documentclass[aps,prd,reprint,superscriptaddress,nofootinbib,amsfonts,amssymb,amsmath]{revtex4-1}
\usepackage[utf8x]{inputenc}
\usepackage{graphicx}
\usepackage{float}
\usepackage{wrapfig}
\usepackage{bm}
\usepackage{color}
\usepackage{comment}
\usepackage{xcolor}
\usepackage[normalem]{ulem}
\usepackage{hyperref}
\usepackage{mathrsfs}
\usepackage{mhchem}
\usepackage{{booktabs}}
\usepackage{natbib}

\hypersetup{
colorlinks=true,
citecolor=blue,
citebordercolor=red,
linktoc=all,
linkcolor=blue,
urlcolor=blue
}
\allowdisplaybreaks

\catcode`\@=11
\def\lsim{\mathrel{\mathpalette\@versim<}}
\def\gsim{\mathrel{\mathpalette\@versim>}}
\def\@versim#1#2{\vcenter{\offinterlineskip
\ialign{$\m@th#1\hfil##\hfil$\crcr#2\crcr\sim\crcr } }}
\catcode`\@=12

\newcommand{\p}{\partial}

\newcommand{\al}[1]{\begin{align}#1\end{align}}
\newcommand{\bp}{\begin{pmatrix}}
\newcommand{\ep}{\end{pmatrix}}
\newcommand{\nn}{\nonumber\\}

\newcommand{\df}{\text{d}}

\newcommand{\bs}[1]{\boldsymbol}

\newcommand{\Tr}{{\rm Tr}\,}

\newbox{\ORCIDicon}
\sbox{\ORCIDicon}{\large\includegraphics[width=0.8em]{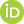}}

\begin{document}

%opening
\title{
Determining wave equations in holographic QCD from Wilsonian renormalization group
}

\author{Fei \surname{Gao}\,\href{https://orcid.org/0000-0001-5925-5110}{\usebox{\ORCIDicon}}\,}
\email{hiei@pku.edu.cn}
\affiliation{Centre for High Energy Physics, Peking University, Beijing 100871, China}

\author{Masatoshi Yamada\,\href{https://orcid.org/0000-0002-2563-9826}{\usebox{\ORCIDicon}}\,}
\email[]{m.yamada@thphys.uni-heidelberg.de}
\affiliation{Institut f\"ur Theoretische Physik, Universit\"at Heidelberg, Philosophenweg 16, 69120 Heidelberg, Germany}

\begin{abstract}
We show a possible way to build the AdS/CFT correspondence starting from the quantum field theory side based on renormalization group approach. An extra dimension is naturally introduced in our scheme as the renomalization scale. The holographic wave equations are derived, with the potential term being determined by the QFT properties. We discover that only around the fixed point, i.e. the conformal limit, the potential in the bulk equations can be fully constrained, and upon this foundation, the correspondence is build. We demonstrate this fact using a 3D scalar theory in which, besides the trivial fixed point, there exists the Wilson-Fisher fixed point. From the energy scalings around those fixed points, we determine the behavior of the potential in the bulk equations.
\end{abstract}

\maketitle

\section{Introduction}
The AdS/CFT (or gauge/gravity) correspondence~\cite{Maldacena:1997re} provides a possible way to understanding nonperturbative nature of   gauge theories. The correspondence conjectures that strongly correlated gauge theories in $d$-dimensional spacetime can be  mapped onto weak gravitational theories in ($d+1$)-dimensional spacetime.

The studies of the AdS/CFT correspondence may include mainly two manners: One is to start from a string theory, choosing the background in such a way as to reproduce essential ingredients of, for instance, quantum chromodynamics (QCD) as matter in the fundamental representation, and to study the resulting QCD-like theories, obtained by top-down approach, e.g., the Witten-Sakai-Sugimoto model, see a review~\cite{Rebhan:2014rxa}. The other,
bottom-up approach is to begin with QCD and attempt
to determine or constrain the dual theory properties by
matching them to known properties of QCD using gauge/gravity correspondence~\cite{Brodsky:2006uqa,deTeramond:2008ht,Brodsky:2014yha,Erlich:2005qh,Karch:2006pv}.

One  important correspondence is that the Regge trajectory for meson mass spectra  in QCD or QCD-like theories, which  is known for instance that the squared masses of excited vector (rho) mesons can be well explained by
$
m_n^2 \sim n\,,
\label{eq: Regge trajectory}
$
with $n$ a consecutive number,  can be achieved via a holographic theory with potential $U(z)\sim z^2$ at $z\rightarrow \infty$.   Hence, the task  now is how such a potential $U(z)\sim z^2$ occurs from analytical computations in quantum field theory (QFT) models, as it is expected that $U(z)$ involves information about the  QFT dynamics. 

The purpose of this work is to propose a way to determine $U(z)$ from the QFT side. All information about the QFT dynamics is encoded in the (one-particle irreducible) effective action $\Gamma$, and can be studied via the functional renormalization group (FRG)~\cite{Wilson:1973jj}. There, the effective average action $\Gamma_k$ is defined such that high momentum modes $|p|>k$ for a cutoff scale $k$ are integrated out. The FRG equation follows the change of $\Gamma_k$ by sliding $k$ as a functional differential equation and one obtains the full effective action as $\Gamma=\Gamma_{k\to 0}$.
In addition to the fact that FRG has been elucidated the nonperturbative nature in QFT (see e.g. Ref.~\cite{Dupuis:2020fhh}), an important feature of the method is the existence of fixed point at which $\Gamma_k$ is independent of $k$, i.e. the system becomes scale invariant. The perturbation of the effective action around the fixed point gives universal critical exponents. Heuristically, the dynamics of  the fixed point  is   related to the conformal property of the theory and presumably also to the dynamics in AdS spacetime. 

In this work, we show that the FRG equation can be reformulated in a form of the wave equation in a holographic theory by a certain change of variables. 
We demonstrate using the 3D scalar field theory that energy scaling at fixed points uniquely determines  the  form of $U(z)$, and  illustrate how the conformal property of QFT is encoded in the anti–de Sitter (AdS) dynamics in this correspondence.

\section{Holographic wave equations from functional renormalization group}
The AdS metric $g_{MN}$ is defined by
\al{
\df s^2 =g_{MN}\df x^M \df x^N = \frac{R^2}{z^2}(\df z^2 + \eta_{\mu\nu} \df x^\mu \df x^\nu)\,,
\label{eq: ads coordinate}
}
where $M,\,N=0,\ldots, d$, while $\mu,\,\nu=0,\ldots, d-1$. Here, $x^M$ is the ($d+1$)-dimensional spacetime coordinate that is decomposed into $d$-dimensional spacetime coordinate $x_\mu$ and an extra dimensional space coordinate $z$. The $d$ dimensional flat metric is denoted by $\eta_{\mu\nu}=\text{diag}(-1,1,\ldots,1)$ and $R$ is the radius of AdS space. 
We consider the equations of motion (the wave equations) for a field $\hat\Phi_J(x,z)\equiv \hat\Phi_{M_1\ldots M_J}(x,z)$ with integer spin $J$ in the AdS spacetime \eqref{eq: ads coordinate}. Assuming that the field $\hat\Phi_J(x,z)$ propagates as a plane wave in the $x_\mu$ directions, namely, $\hat\Phi_J(x,z)=e^{iP_\mu x^\mu-\varphi(z)/2}\Phi_J(z)$, the wave equations for $\Phi_J(z)$ can then read as
\al{
\left[\p_z^2 - \frac{(d-1-2J)}{z}\p_z - \frac{(\mu R)^2}{z^2}- U_J(z) + {\mathcal M}^2\right]\Phi_J(z)=0 \,,
\label{eq: bulk wave equation}
}
where $\mu^2$ is the squared mass parameter of $\Phi_J$ and is given by $(\mu R)^2=(\Delta-J)(\Delta-d+J)$ according to Refs.~\cite{lYi:1998trg,Brodsky:2014yha} with conformal dimension $\Delta$, and $\mathcal M^2=P_\mu P^\mu=\eta^{\mu\nu}P_\mu P_\nu$ is the invariant mass.
Here, $\varphi(z)$ is the dilaton field that constitutes the potential
\al{
U_J(z)=\frac{1}{2}\varphi'' +\frac{1}{4}\varphi'^2+\frac{2J-d+1}{2z}\varphi'\,,
\label{eq: potential from dilaton field}
}
in which the prime denotes the derivative with respect to $z$.
The explicit derivation of Eq.~\eqref{eq: bulk wave equation} is presented, e.g., in Ref.~\cite{Brodsky:2014yha}.

Through redefining the field $\Phi_J(z)$ as $\psi(z)=(R/z)^{-\frac{d-1-2J}{2}}\Phi_J(z)$, Eq.~\eqref{eq: bulk wave equation} can be written in terms of the Schr\"odinger equation for $\psi$ on the bulk:
\al{
\left( -\frac{\df^2}{\df z^2} -\frac{1-4L^2}{4z^2} + U_J(z) \right)\psi(z) ={\mathcal M}^2\psi(z)\,,
\label{eq: Schrodinger equation}
}
where $L^2$ is the Casimir representation of orbital angular momentum and $L^2=(J-d/2)^2+(\mu R)^2$. The equivalence between Eqs.~\eqref{eq: bulk wave equation} and \eqref{eq: Schrodinger equation} has been shown in Ref.~\cite{deTeramond:2008ht}. 
If $U(z) \sim z^2$ for a large $z$, then its solution reproduces discrete spectra ${\mathcal M}^2=P^2\sim n$ as eigenvalues~\cite{Karch:2006pv}.
From Eq.~\eqref{eq: potential from dilaton field}, the dilaton field behaving as $\varphi(z)= \lambda z^2$ explicitly provides $U_J(z)=\lambda^2z^2 + 2\lambda(J-1)$ with $\lambda$ as a constant parameter.
Elucidating the origin of such a potential or a dilaton field is an important issue in holographic QCD.

We show from now that Eq.~\eqref{eq: bulk wave equation} can be derived from the QFT language and $U_J(z)$ is related to the beta function.
To this end, we start by introducing the FRG equation (or simply the flow equation). One of its forms is formulated for the one-particle irreducible effective action $\Gamma_k$ in $d$-dimensional spacetime and is given by~\cite{Wetterich:1992yh}:
\al{
k\p_{k} \Gamma_k =\frac{1}{2} \Tr \left[ k\p_{k} R_k(p)\cdot G_k(p)\right]\equiv \beta_\Gamma\,,
\label{eq: FRG equation}
}
where $k$ is the energy scale, $\p_{k}=\p/\p k$ and Tr is the functional trace acting on all internal spaces in which field variables are defined. Here $R_k(p)$ is the regulator function to realize the coarse-graining procedure, and $G_k(p)$ is the regulated full propagator whose explicit form is given by $G_k(p) =(\Gamma_k^{(2)}(p)+ R_k(p))^{-1}$
with the full two-point function $\Gamma_k^{(2)}$, i.e. the second-order functional derivative with respect to field variables.
The $n$-point function is defined by the functional derivatives 
\al{
&\left[\Gamma_k^{(n)}(p_1,\cdots,p_n)\right]_{\{M\}\cdots \{N\}}\nn
&\qquad=\frac{\delta^n\Gamma_k}{\delta \phi_{M_1\ldots M_i}(p_1)\cdots\delta \phi_{N_1\ldots N_j}(p_n)},
\label{eq: n point function}
}
where $\{M\}=M_1\ldots M_i$ stands for a set of momenta and one of momenta is redundant thanks to the momentum conservation.
Note that $\Gamma^{(n)}_k$ with different momenta $p$ together with, e.g., the Lorentz structures correspond to different channels of the wave equation. Each can be selected by choosing the momentum and the projection.

The flow equation for the full $n$-point function $\Gamma_k^{(n)}$ can be obtained by taking the $n$ th order functional derivative for both sides of Eq.~\eqref{eq: FRG equation} with respect to field variables.
More specifically, we symbolically write
\al{
k\p_k\Gamma_k^{(n)} = \frac{\delta^n\beta_\Gamma}{\delta \phi_{J_i}(p_1)\cdots\delta \phi_{J_j}(p_n)} \equiv \beta_\Gamma^{(n)}.
\label{eq: flow equation for n point function}
}
where $\phi_{J_i}(p)\equiv \phi_{M_1\ldots M_i}(p)$.
The regulator function satisfies $R_k(p)>0$ for $p^2/k^2\to 0$ and behaves as $R_k(p)= 0$ for $k^2/p^2\to 0$ and as $R_k(p)\to \infty$ for $k^2\to\infty$. There is an infinite number of possible forms for such a function. A specific choice of its form corresponds to the renormalization scheme.

An important fact is that the functional renormalization group equation \eqref{eq: FRG equation} is derived without any approximations. In other words, solving the flow equation without making approximations means exactly performing the path integral and obtaining full information about the QFT dynamics.

Now we show that the flow equation can be written in the form of the bulk wave equation. 
A similar attempt to build relations between these two conceptions has been made in Refs.~\cite{Heemskerk:2010hk,Sathiapalan:2017frk,Sathiapalan:2019zex,Sathiapalan:2020cff,Dharanipragada:2022iud,Kim:2021qbi,Guijosa:2022jdo}. 
See also Refs.~\cite{Basile:2021euh,Basile:2021krk,Basile:2021krr} as applications of the FRG to the bosonic string dynamics.
The previous studies mostly focus on  the holographic renormalization group in AdS spacetime and the correspondence is build upon its similarity to the Wilsonian RG in QFT, which is the RG applied here. Such a comparison is instructive; however, as raised in Ref.~\cite{Heemskerk:2010hk}, a key question, ``what cutoff on the field theory corresponds to a radial cutoff in the bulk?'', is left unanswered. In this work, instead of applying the holographic renormalization group, we focus on the Wilsonian RG equation for the $n$-point correlation function $\Gamma_k^{(n)}$ and shows the exact correspondence to the holographic wave equations \eqref{eq: bulk wave equation}. The comparison reveals the importance of the fixed point dynamics, which naturally provides an answer about the cutoff.

We start by multiplying $k^{-\delta}$ with a power factor $\delta$ and then taking a derivative with respect to $k\p_k$ on the flow equation \eqref{eq: flow equation for n point function} and rewrite the flow equation as 
\al{
k\p_{k}\left\{ k^{-\delta} k\p_{k}\Gamma^{(n)}_k\right\} =k\p_{k}\left( k^{-\delta}  \beta^{(n)}_\Gamma \right)\,.
\label{eq: FRG equation derivatived}
}
Then, we perform the following change of variables:
\al{
k=1/z,
}
and
\al{
%&\Gamma^{(n)}_k(p_1,\ldots, p_n)= [z^{d-\eta}\Phi_J(z)]^{-1}\,,
&\left[\Gamma^{(n)}_k(p_1,\ldots,p_{i-1};p_i,\ldots,p_{n})\right]_{\{M\}}\nn
&=\delta^d(p_1+\cdots+p_{i-1}-p_i-\ldots-p_{n}) T_i^{\{M\}}[z^{d-\eta}\Phi_J(z)]^{-1},
\label{eq:n point function for bound state}
%\\
%&\Phi_J(z)=\int \df^dx\, e^{iP\cdot x z} \phi(x)
}
where $\eta$ is a constant and can be related to conformal dimension $\Delta$,  and spin $J$ as will show later in Eq.~\eqref{eq:n point function for bound state}. 
The correlation function in \eqref{eq:n point function for bound state} has been chosen to contain the bound state channel as  $T_n^{\{M\}}$ an appropriate tensor structure for the symmetries of the system in the channel that projects out the respective bound states  and the momentum parametrization as $p_1+\cdots+p_{i-1}=P$ with $P^2= {\mathcal M}^2$.   Note that there are in fact arbitrarily many different correlation functions which can contain the same bound state, and hence there are different choices of momentum and Lorentz structure in   Eq.~\eqref{eq:n point function for bound state}. In principle, the obtained spectra of the bound state should be equivalent.   The only artificial dependence in  $\Phi_J(z)$ comes from the    regularization $R_k(p)$ of the correlation function $\Gamma^{(n)}_k$.  The dependence of regularization $R_k(p)$ does not change the equations as it can be absorbed into the definition of wave function $\Phi_J$.  However, such dependence in $\Phi_J$  is required to  be vanishing for the correspondence between Eqs.~\eqref{eq: bulk wave equation} and \eqref{eq: FRG equation derivatived} since the spectra of the bound states cannot depend on the regularization.  The key point of our paper is to show that this is only valid in the critical region near fixed point, and  we will discuss this in detail in the next section.

For the left-hand side of the flow equation \eqref{eq: FRG equation derivatived}, one finds

\begin{widetext}
\al{
\text{(LHS of Eq.~\eqref{eq: FRG equation derivatived})}=-z^{\delta-\eta+d+2}\left(\Gamma_k^{(n)}\right)^2\left(
\p_z^2 -  \frac{( -1-\delta + 2d-2\eta)}{z}\p_z  -2 \left[\frac{\p\log \Phi_J}{\p z}\right]^2 - \frac{(d-\eta-\delta)(d-\eta)}{z^2}
  \right) \Phi_\eta(z)\,.\label{eq: lhs}
}
\end{widetext}

From the right-hand side of the flow equation \eqref{eq: FRG equation derivatived}, one can recognize the  potential $U(z)$ as

\al{
U_J(z) = -\frac{z^{-1-\delta}}{\Gamma_{z}^{(n)}}\p_{z}\left( z^\delta  \beta^{(n)}_{\Gamma_k} \right),
\label{eq: identification of potential}
}
where $\Gamma_k^{(n)}=\Gamma_{1/z}^{(n)}\equiv \Gamma_z^{(n)}$. Note that $ \beta^{(n)}_{\Gamma_k}= -\beta^{(n)}_{\Gamma_z}=-z\p_z\Gamma_z^{(n)}$.
Together with the identification of variables as
\al{
&\delta=2J+d-2\eta\,,&
& {\mathcal M}^2 = -2\left[\frac{\p \log \Phi_J}{\p z}\right]^2 \,,
\label{eq: redefinition of variables}
}
with the parameter $\eta$  being related to  conformal dimension $\Delta$ and spin $J$ as
\al{
(\eta-J-d/2)^2=2(J-d/2)^2-(\Delta-d/2)^2\,.
\label{eq: angular momentum part}
}
After this we can see that the flow equation \eqref{eq: FRG equation derivatived} is equivalent to the bulk wave equations~\eqref{eq: bulk wave equation}. We note here that the last term in Eq.~\eqref{eq: lhs} with the redefinition \eqref{eq: redefinition of variables} is 
\al{
(\mu R)^2&=(d-\eta-\delta)(d-\eta)\nn
&= (J-d/2)^2-\delta^2/4\,.
\label{eq: angular momentum part}
}
This quantity corresponds to the angular momentum contribution, and eventually leads to $L^2=(J-d/2)^2+(\mu R)^2=2(J-d/2)^2-\delta^2/4$ in  Eq.~\eqref{eq: Schrodinger equation}. If $L^2\geq0$, then one has $(\mu R)^2\geq-(J-d/2)^2\sim-4$ with $J=0$ and $d=4$, which is the equivalence between the quantum mechanical stability condition and the Breitenlohner-Freedman stability bound in AdS space similarly as argued in Ref.~\cite{deTeramond:2008ht}.

Before discussing the evaluation of $U(z)$ within a specific model, we briefly review the notion of fixed points and scaling (critical) exponents in the functional renormalization group.
Fixed points characterize the scale invariance of a theory, i.e., certain points so as to be $\beta_\Gamma=0$. Let us here denote $\Gamma_*$ a scale invariant action. For such a fixed point, consider a small perturbation $\delta\Gamma_k=\Gamma_k-\Gamma_*$ around $\Gamma_*$. 
The perturbation also works on any $n$-point function through Eq.~\eqref{eq: n point function}. 
Now we generally expand the $i$-point function  as
\al{
\left(\Gamma^{{(i)}}[p_1,\ldots,p_i;\bar\phi]\right)^{M_1\cdots M_i} = g^i_{k}(p_1,\cdots,p_i) T_i^{M_1\cdots M_i} \,,
\label{eq:fixed point expansion}
}
where $g_k^i$ are couplings depending on $k$ and external momenta. We note that the external momenta should be in principle arranged to contain an on shell bound state, but at the lowest order in the derivative expansion of the effective action can also be taken in vanishing external momenta for $\tilde g_k^i(0,\ldots,0)$.
Let $d_i$ be the  mass dimension of $g_k^i$. The flow equation \eqref{eq: flow equation for n point function} is reduced to a coupled differential equation for the dimensionless couplings $\bar g_k^i=g_k^i/k^{d_i}$, i.e., $k\p_{k}\bar g_k^i =\beta_{i}(\{\bar g_k\})$ where $\{\bar g_k\}$ stands for a set of couplings.
Within the operator expansion scheme, a fixed action $\Gamma_*$ infers vanishing beta functions $\beta_i=0$ for all $i$ which gives the set of fixed couplings $\{\bar g_{k*}^{i}\}$ and the small perturbation of the $i$-point function $\delta\Gamma^i_k$ is given in terms of couplings:
\al{
\bar g_k^i = \bar g_{k*}^{i} + \sum_i C_i^j\left( \frac{k_0}{k} \right)^{\theta_j} \,,
\label{eq: RG flow around FP}
}
where $k_0$ is a reference scale and $C_i^j$ is a matrix representing mixing effects between different couplings $i\neq j$. 
When the mixing effects are negligible, one has $C_i^j\approx c_i\delta_i^j$ with $c_i$ small parameters.
Here, $\theta_j$ are associated critical exponents characterizing the energy scalings of operators around the fixed point. 
Couplings with a positive scaling exponent are amplified for $k\to 0$ and then are called ``relevant," while ``irrelevant" couplings have a negative $\theta_i$ and thus decrease for the IR limit. In particular, at the trivial (or Gaussian) fixed point $\bar g_{k*}^i=0$, the critical exponents correspond to the canonical mass dimension of coupling constants, i.e., $\theta_i=d-d_i$.

\section{Relation between critical exponent and bound state nature}
We see from the derivation of the  wave equation presented above that the  dynamics in AdS spacetime can be fully captured by the FRG equation, with all information about the QFT dynamics in $d$-dimensional spacetime now being included in $U(z)$. 
The interesting question now is what   the potential $U(z)$ looks like  from the functional renormalization group analysis, i.e. directly evaluating the right-hand side of Eq.~\eqref{eq: identification of potential}. 
Before discussing the behavior of potential,  some general conclusions can be drawn through the above derivation:

\begin{itemize}
\item[(i)] The conformal property of QFT is encoded in AdS dynamics in AdS/CFT correspondence through the fixed point of the theory.
\end{itemize}

As mentioned above, in the FRG equation, there is generally the regularization $R_k(p)$ dependence. This regularization term is an arbitrary function in $k$, with only constraint at $k=0$ to return to the original theory.  Such an artificial regularization term makes the potential $U(z=1/k)$  underdetermined, and consequently  the bound state  mass  ${\mathcal M}^2$ is  underdetermined as well, with dependence of the regularization term.  

The fixed point is thus important as it ensures the expansion of the correlation functions in QFT as in Eq.~\eqref{eq: RG flow around FP}. The expansion  coefficients $C^j_i$ and the critical exponents $\theta_j$ are independent of the regularization term inside the critical region~\cite{Berges:2000ew} and, hence,  makes the  corresponding potential  uniquely determined. In a word, only around the fixed point, the flow equation \eqref{eq: FRG equation derivatived}  is fully equivalent to the bulk wave equation \eqref{eq: bulk wave equation} with the vanishing of the  dependence of the artificial  regularization term $R_k(p)$.  

It is thus clear that,  if  the mass term ${\mathcal M}^2$ in the bulk wave equation corresponds to the bound states of some theory, there must exist  fixed point in the respective  theory to entail the correspondence, otherwise the bulk wave equation is rather a coincidence to describe the bound states. Moreover,  the question raised in Ref.~\cite{Heemskerk:2010hk} is now naturally answered: the radial cutoff in the bulk is the boundary of the critical region where  the correlation functions of QFT start to deviate from the critical behavior as in the expansion of  Eq.~\eqref{eq: RG flow around FP}.   Note that the potential is vanishing at fixed point which corresponds to the conformal limit, and the critical behavior  approaching fixed point   together with the cutoff characterized  by the critical region determine the mass spectrum of the bound states.

\begin{itemize}
\item[(ii)] For each correlation function  $\Gamma^{(n)}_k$  in a specific channel,  there corresponds to a type of  potential in the bulk which behavior can be determined by the expansion coefficients and  critical exponents of the respective correlation function.
\end{itemize}

 The potential   for the bound states is  determined by the correlation functions of $\Gamma^{(n)}_k$ that contain the respective bound states. Putting the expansion in \eqref{eq:fixed point expansion} and \eqref{eq: RG flow around FP} into the definition of potential in \eqref{eq: identification of potential}, one immediately obtains:

\al{
U(z) =z^{-1-\delta}\frac{\p_{z} (z^{\delta+1} \p_z \bar g_k^i)}{\bar g_k^i}
}

 If setting $\delta=d_i$, then one can cancel the naive contribution from the dimension of the correlation functions and simply has
\al{
U(z)=\frac{ \p_z^2 \bar g_{k}^{i} }{ \bar g_{k}^{i} }
}

Note that the other choices of $\delta$ will just shift the definition of $L^2$. 
Therefore, the parameter $\delta$ should be fixed by relating $L^2$ to the physical angular momentum of the bound states which contain in the respective channel of the $n$-point correlation functions. This fact provides the meaning of $L^2$ in the AdS bulk equations.
Now with all this knowledge, one can see that  the analysis of the potential $U(z)$ can be carried on by analyzing the fixed point of FRG equations.

\section{Bound states in 3D Scalar theory and  gauge theory}
We have shown that the holographic wave function can be derived from the functional renormalization group equation, and how the bound states property is related to the fixed point dynamics. Now we would like to illustrate some special cases in 3D scalar theory and gauge theory.   

\subsection{At trivial fixed point}
For a theory with the asymptotically free property, the trivial fixed point is located at the UV scale (corresponding to $z=0$) while for the others are at $z=\infty$. Around the trivial fixed point, the system can be analyzed perturbatively.

For instance,  the flow equation for the mass parameter in scalar theory is
\al{
&m^2(k) =k^2\bar m^2(k)\,,\nn
&\bar m^2(k) = \bar m_*^2 +\bar m_0^2(k_0/k)^{\theta_m}\sim  \bar m_*^2 + c_m z^{\theta_m}\,,
}
where $m_0^2$ is a constant given at $k_0$.
For the trivial fixed point $\bar m_*^2=\lambda_*=\cdots=0$, one has   $\theta_m= 2$ for which the potential $U(z)$ around the trivial fixed point is found to be
\al{
U(z)=z^{-1-\delta}\frac{\p_{z} (z^{\delta+1} \p_z  m^2)}{m^2} =  \frac{(\theta_m-2)(\delta+\theta_m-2)}{z^2}\,,
\label{eq: perturbative potential}
}
where we have used the trivial fixed point $\bar m_*^2 =0$. The potential at the fixed point with $\theta_m=2$ is vanishing.
Now considering a perturbation around the trivial fixed point that gives a small deviation from $\theta_m=2$ to $\theta_m = 2- a_m$, one obtains $U(z)=- a_m(\delta-a_m)/z^2$.

 The similar derivation can be  done in any QFT theory, and thus one may have $U(z)=- a_m(\delta-a_m)/z^2$ with $a_m$ the anomalous dimension, which again coincides with the angular momentum part \eqref{eq: angular momentum part} in the wave equation is corrected so as to be
\al{-\frac{1-4L^2}{4}\rightarrow -\frac{1-4L^2}{4}-a_m(\delta-a_m)\,.}
This in general shows how the anomalous dimensions  are introduced into the AdS dynamics.

\subsection{At nontrivial fixed point}
For the 3D scalar theory, there exists the Wilson-Fisher fixed point being an IR fixed point and there is only one relevant coupling that gives a nontrivial energy scaling behavior for all correlations. For the two point correlation function  with  $\delta=d_i=2$  we have
\al{
U(z) =\frac{\p_{z} ^2  (\bar m_*^2  +c_m z^{\theta_m})}{\bar m_*^2  +c_m z^{\theta_m}} \sim z^{-2+\theta_m}\,,
\label{eq: non-trivial potential}
}
where we have assumed that $c_m$ is a small perturbative parameter around the nontrivial fixed point and have taken the lowest order of the expansion in terms of $c_m$.  
The elaborated studies on the determination of the scaling exponent   have shown that $\theta_m \sim 1.59$ ~\cite{Balog:2019rrg,Baldazzi:2021ydj}. 
The Schr\"odinger equation with the potential $\sim 1/z^{2-\theta_m}$ then entails some unusual bound states. The numerical computations actually have suggested some bound states exist in 3D $\phi^4$ theory~\cite{Caselle:2000yx,Rose:2016wqz,Horak:2020eng}, which coincides with what we found here.  The additional bound states can survive between $d=2$ to $d=4$ as the interval of the dimension for occurring the  nontrivial fixed point, which  
also verifies the numerical computation~\cite{Rose:2016wqz}.

To demonstrate our approach, we show a few eigenvalues by solving the Schr\"odinger equation \eqref{eq: Schrodinger equation} with the potential \eqref{eq: non-trivial potential} for $J=0$ which yields $L^2=7/2$ in $d=3$. 
To do this, we need to fix two parameters. One of them is the coefficient, denoted here by $\gamma$, in the potential, i.e. $U(z)= \gamma z^{2-\theta_m}$. Although we cannot determine the exact value of $\gamma$, we here would infer from the typical order of the fixed point value $m_*^2\sim {\mathcal O}(1)$ and a small perturbation parameter $c_m$ that $\gamma$ is of order of $0.1$. 
The potential \eqref{eq: non-trivial potential} is valid around the fixed point. Therefore, there must be a cutoff scale $z_\text{cut}= 1/k_\text{cut}$ where the mass parameter starts to deviate from the scaling regime with $\theta_m=1.59$ under the RG evolution.  In this study, we set $\gamma=-0.15$ and similarly to the hard wall model~\cite{Polchinski:2001tt,Karch:2006pv}, we set $z_\text{cut}=10$ and solve the Schr\"odinger equation with the boundary condition $\psi(z_\text{cut})=0$ .

\begin{figure}
\includegraphics[width=8cm]{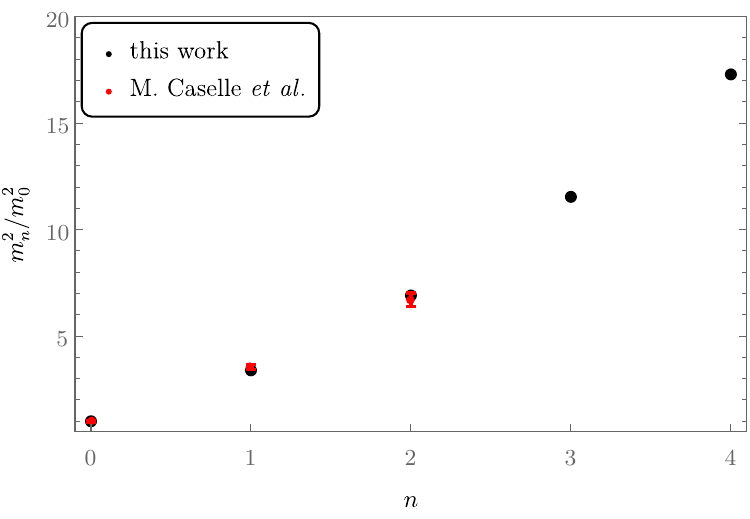}
\caption{
Excited mass eigenvalues of bound states with $J=0$ in a $3D$ scalar field theory.
Black points show ratios between eigenvalues of ground and excited states as solutions to the Schr\"odinger equation with $U(z)=-0.15z^{-0.41}$ and the boundary condition $\psi(z_\text{cut}=10)=0$. Red points with error bar are taken from Ref.~\cite{Caselle:2000yx}.
}
\label{fig:eigenvalues} 
\end{figure}

We compare energy eigenvalues obtained between in Ref.~\cite{Caselle:2000yx} and in our work. In order to avoid mismatches of the energy scale, the ratios between excited states and the ground state are shown in Fig.~\ref{fig:eigenvalues}. It seems that our approach captures the property of bound states in $3D$ scalar field theory despite the simple and crude calculation. We also give our prediction for the higher excited states ($n=3,\,4$). These mass eigenvalues should be tested by other methods.
More accurate FRG computations could provide precise values for $\gamma$ and $z_\text{cut}$ without any ambiguities. We would leave such an analysis for future works.

Now, it is especially interesting to study the nontrivial fixed point for gauge theory. In particular,  it is expected that for QCD, there is presumably an infrared fixed point~\cite{Ojima:1978hy,Kugo:1979gm,Gies:2002af,Pawlowski:2003hq,Aguilar:2002tc}.   This fixed point and strong coupling nature of QCD lead to a large scaling window for studying hadrons, which is the standing point of the holographic QCD~\cite{Brodsky:2010ur,Deur:2016tte}.  However, due to the complexity of the gauge theory, here we will not do the computation directly but instead, we  draw some conclusions from the correspondence. 

Based on the derivation in this work, we can confirm that the infrared fixed point should  exist in order to reproduce the potential $U(z)\sim z^2$ at $z\rightarrow \infty$, and consequently, the Regge trajectory. In another word, the Regge trajectory reveals that in QCD there must exist an infrared fixed point at which an associate critical exponent reproduces $U(z)\sim z^2$ via Eq.~\eqref{eq: identification of potential}.  More interestingly, the $z^2$ dependence of the potential in turn leads:
\al{
[\Phi]^{-1}\sim \left(g_{\mu\nu}-\frac{P_\mu P_\nu}{P^2} \right)\frac{\delta^2  \Gamma_k}{\delta J_\mu\delta J_\nu}\sim\frac{1}{k^4}\,,
}
with $J_\mu\sim\bar{\psi}\gamma_\mu\psi$  the vector meson field or in QCD, the vector current of quarks, respectively.  The relation implies  a linear confining potential between the vector current of quarks, which is associated with the Regge trajectory.

\section{Summary}
In this work, we show that the holographic wave equation can be obtained from the quantum field theory via the renormalization group method.
One may start from the flow equation and then obtain the potential $U(z)$, the metric and the dilaton background introduced in the AdS space are then fixed. In this sense,  the flow equation describes the dynamics in different AdS spaces by  equipping  with different fixed points which  characterize conformalities of the system. This picture seems natural.
The additional dimension in the AdS spacetime is naturally interpreted as the renormalization scale, and the conformality is encoded through the fixed point of the  renormalization group flow, to guarantee the uniqueness of the correspondence. 

Practically, we show  that  the anomalous dimensions can be  introduced in the AdS dynamics through the trivial fixed point in perturbative region. After that,  we verify that the Wilson-Fisher fixed point brings in  some  bound states in $3D$ scalar theory as discovered  from some numerical studies.  Due to the complexity of QCD dynamics, it is hard to directly compute the potential $U(z)$ from QCD. Nevertheless, what we discovered here is that the Regge trajectory naturally entails a nontrivial fixed point to occur  in the infrared of QCD.  In turn, the critical exponent at this fixed point uniquely determines the correspondence of AdS spacetime and greatly simplifies the low-energy QCD dynamics.

\section*{Acknowledgements}
We thank J.~M.~Pawlowski and A.~Pastor-Guti\'errez for valuable discussions. F.\,G. also thanks all the other members in fQCD collaboration~\cite{fQCD:2022}. 
We are supported by the Alexander von Humboldt Foundation.
The work of M.\,Y. is also supported by the DFG Collaborative Research Centre ``SFB 1225 (ISOQUANT)" and Germany’s Excellence Strategy Grant No.~EXC-2181/1-390900948 (the Heidelberg Excellence Cluster STRUCTURES).

\bibliography{refs.bib}
\end{document}